\documentclass{iauc}
\usepackage{graphicx}
\usepackage{psfig}

\title[Rings around extrasolar planets] 
{Detectability of planetary rings around an extrasolar planet from reflected-light photometry }

\author[L. Arnold \& J. Schneider]   
{L. Arnold$^1$%
~\and J. Schneider$^2$}

\affiliation{$^1$CNRS - Observatoire de Haute-Provence 04870 Saint-Michel-l'Observatoire, France \break 
email: arnold at obs-hp.fr\\[\affilskip]
$^2$Observatoire de Paris-Meudon, 92195, Meudon Cedex, France \break email: Jean.Schneider at obspm.fr}

\pubyear{2005}
\volume{200} 
\pagerange{119--126}
\date{?? and in revised form ??}
\jname{Direct Imaging of Exoplanets: Science and Techniques}
\editors{Claude Aime, Farrokh Vakili, eds.}
\begin{document}

\maketitle

\begin{abstract}
The next generation of high-contrast imaging instruments will provide the first unresolved image of an extrasolar planet. While the emitted infrared light from the planet in thermal equilibrium should show almost
no phase effect, the reflected visible light will vary with the orbital phase angle. We study the photometric variation of the reflected light with orbital phase of a ringed extrasolar planet. We show that a ring around an extrasolar planet, both obviously unresolved, can be detected by its specific photometric signature.
\end{abstract}

\firstsection 
\section{Introduction}

The discovery of extrasolar planets by radial velocity measurements has provided the first dynamical characteristics of planets (orbital
elements and mass). The next step will be to investigate physical characteristics of extrasolar (albedo, temperature, radius etc.) and their
surroundings. Among the latter are planetary rings. Coming space or ground-based high-contrast imaging instruments will be able to 
provide the first image of an unresolved extrasolar planet, in the thermal infrared or in the visible light. 
The emitted thermal infrared light from the planet should show no phase effect assuming the planet is in thermal equilibrium. 
But the reflected visible light will vary with phase angle, as should be shown by a broad-band photometric follow-up of the planet during its orbital motion.

We thus argue that it is of interest to study how the presence of a ring around a planet would influence its
brightness as a function of its orbital position. This paper shows a few examples
 (more in \cite{arnold04}) of how the reflected light curve of a ringed planet is different from that of a ringless planet, thus revealing the presence of the ring. The specific signature of a ring has been already mentioned and briefly qualitatively discussed (\cite{schneider01}), but here a more quantitative model is considered, in which the basic optical properties of the
planet and the ring are taken into account, together with all geometrical parameters describing the ringed planet.

\section{On the relevance of the existence of ringed planets}

Although all giant planets of the Solar System have rings, ringlets or arclets, those of
Saturn are by far the brightest. But at methane absorption wavelengths, giants planets are fainter than in the visible and their
rings appear relatively brighter, increasing their detectability. For Uranus for example, the rings appear with a brightness 
comparable to the planet (\cite{lellouch_et_al2002}). 
Even the Earth might have been surrounded by a ring, but only during $10^5$ to several $10^6$ years (see \cite{arnold04} for a detailed bibliography). 
Although Saturn rings are probably younger than the planet itself, $10^8$ years (\cite{cuzzy98}), i.e. one order
of magnitude younger than the planet, we consider that it remains relevant to look for rings around extrasolar planets, at least for giant planets.

\section{The ringed planet model}
\label{section_model} 

To compute the light curve of the ringed planet as a function of the orbital phase angle, we build radiance maps of the
object, for a given set of geometrical parameters (\cite{arnold04}). We assume that the planet is an isotropic (lambertian) gray spherical
light diffuser.
The ring is a planar, homogeneous and anisotropic (non-lambertian) gray scattering layer. The ring brightness is estimated by assuming only single scattering in the ring.
Mutual lighting (i.e. planet-shine on the ring, or ring-shine on the planet) is not modeled, but mutual shadowing is computed quite easily. All curves are normalized to the flux of a ringless planet seen in quadrature.

\section{Discussion}
\label{section_discuss} 
\subsection{A ringed planet observed at inclination $i=0^{\circ}$}
\begin{figure}
 	\includegraphics[width=6.7cm]{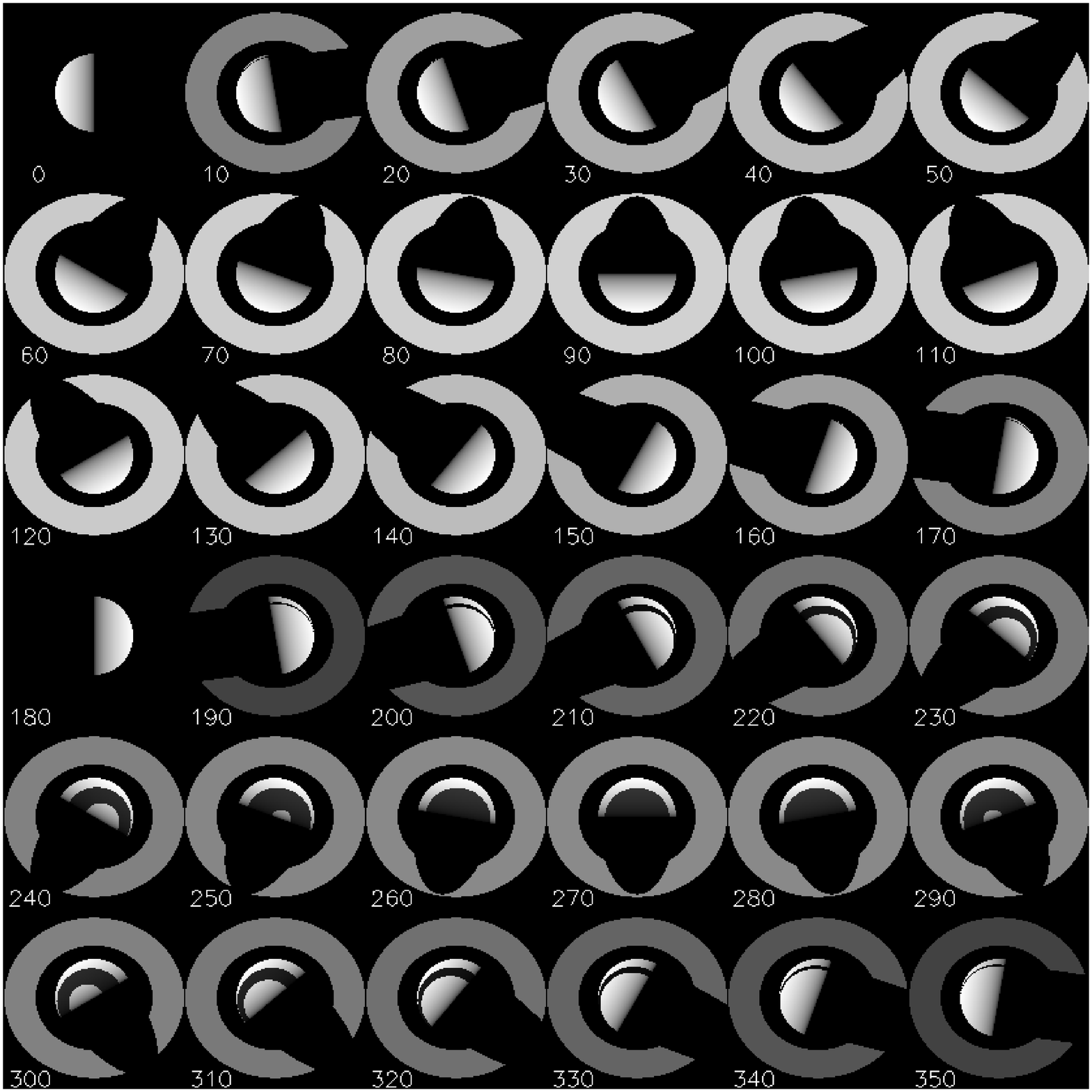}
 	\includegraphics[width=6.7cm]{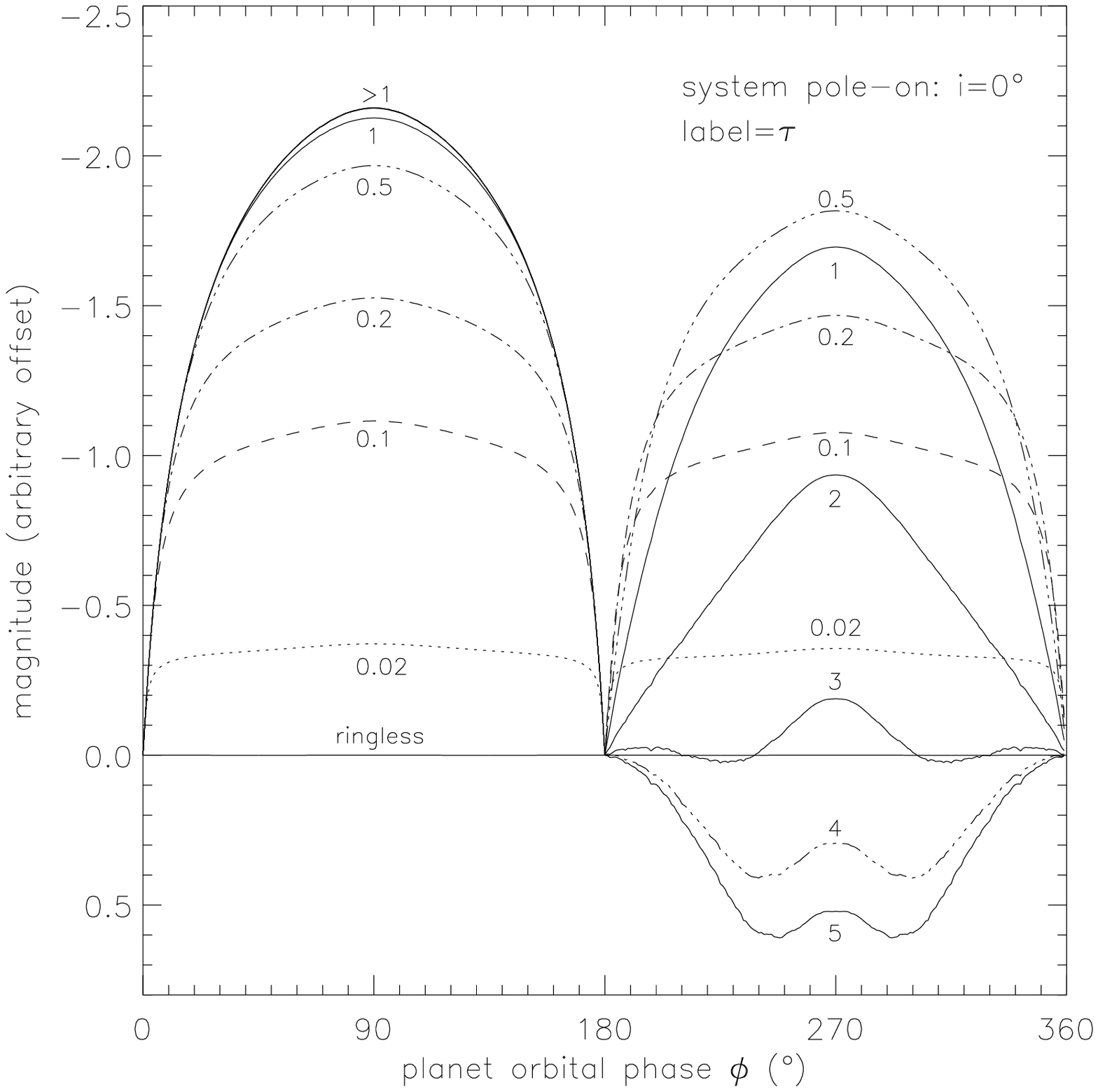}
  \caption{Extrasolar system observed pole-on at different orbital positions (labels, image scale = radiance$^{0.33}$). 
  Inclination $i=0^{\circ}$, the planet is always seen in quadrature, while the ring, with a Saturn-like obliquity of $i_r=26.73^{\circ}$, shows its 
  illuminated face during the first half-orbit. During the second half-orbit, the back-illuminated ring becomes fainter. Corresponding light curves 
  are given at right for different optical thicknesses $\tau$ of the rings (labels).}
  \label{fig01}
\end{figure}

Let's first consider the simple geometrical configuration where an extrasolar system is observed pole-on. This configuration means that the planet always remains in quadrature, thus showing always the same phase. For a ringless planet, it means that its brightness remains constant during the full orbit (assumed circular). Now for
a ringed planet with non-zero obliquity, the ring alternately shows its illuminated and dark side, moreover it projects its shadow on the planet. This is illustrated in figure \ref{fig01} showing strong slope changes at the equinoxes. This dichotomy of the light curves is due to the ring being alternately seen in reflection and in transmission. This is a clear ring signature which is often well visible in other geometrical configurations.  

\subsection{Dual-band photometry of a ringed planet}
\begin{figure}
   \includegraphics[width=6.7cm]{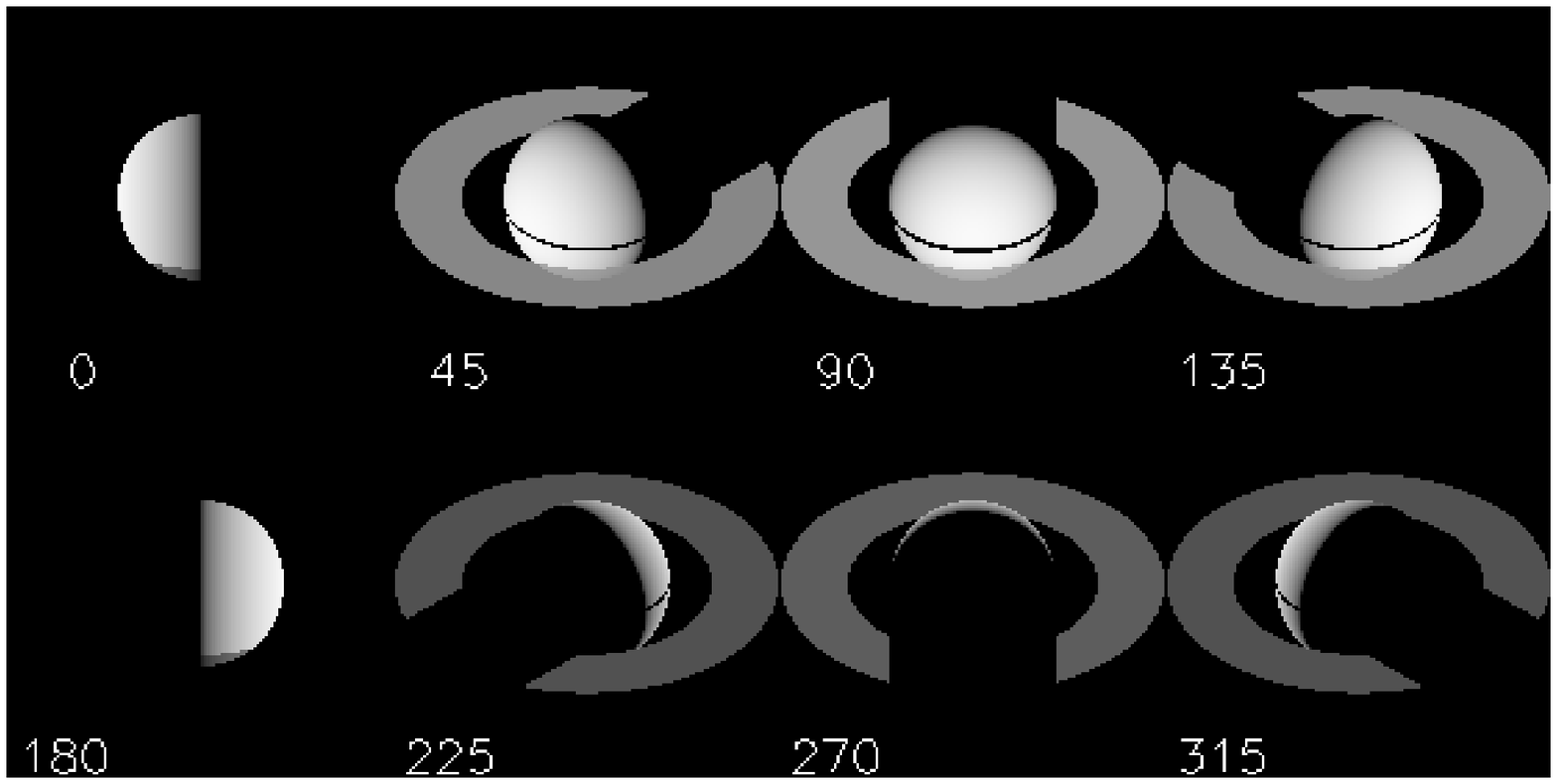}\\
   \vskip-3.7cm
   \includegraphics[width=6.7cm]{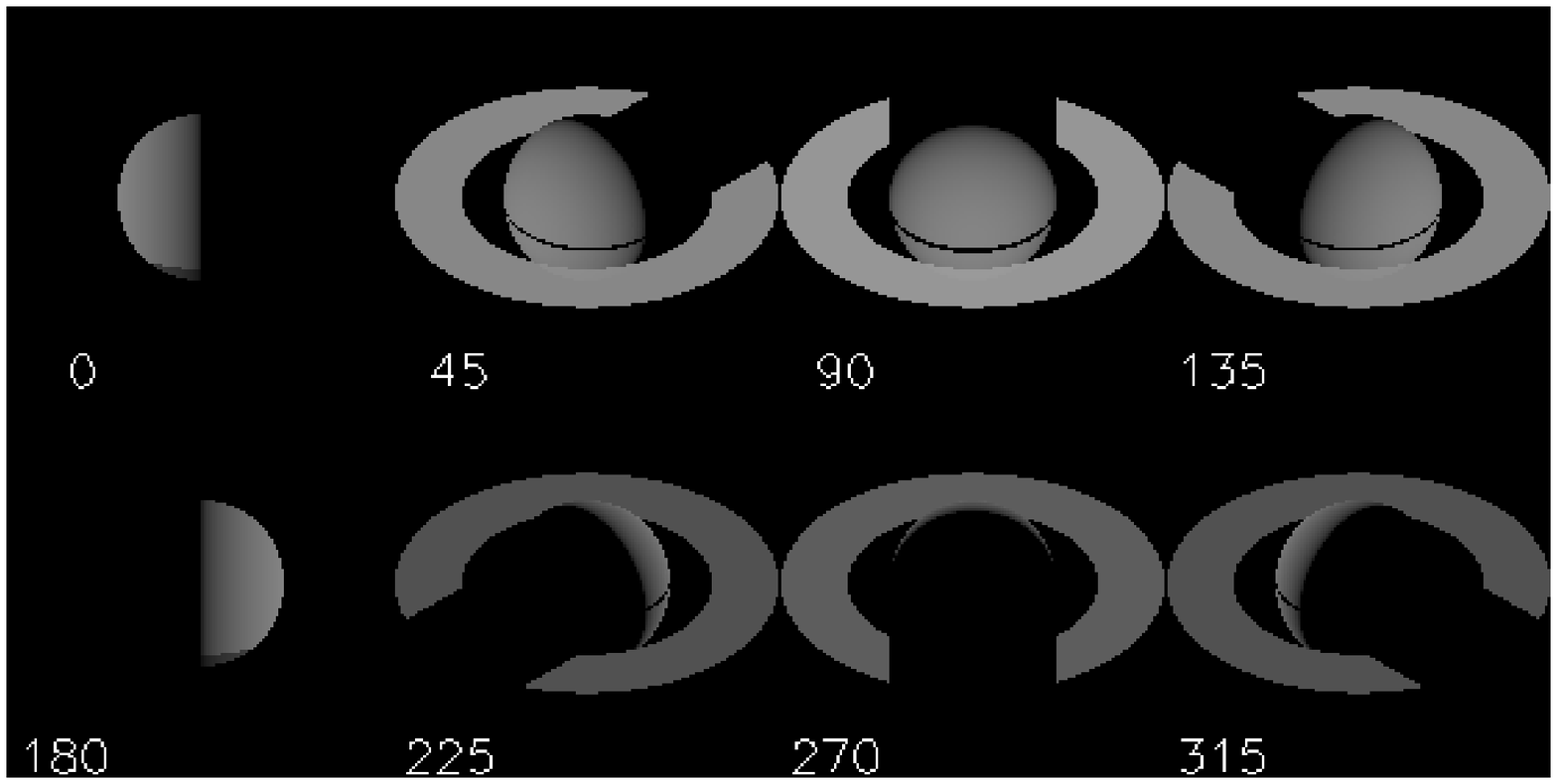}
   \includegraphics[width=6.7cm]{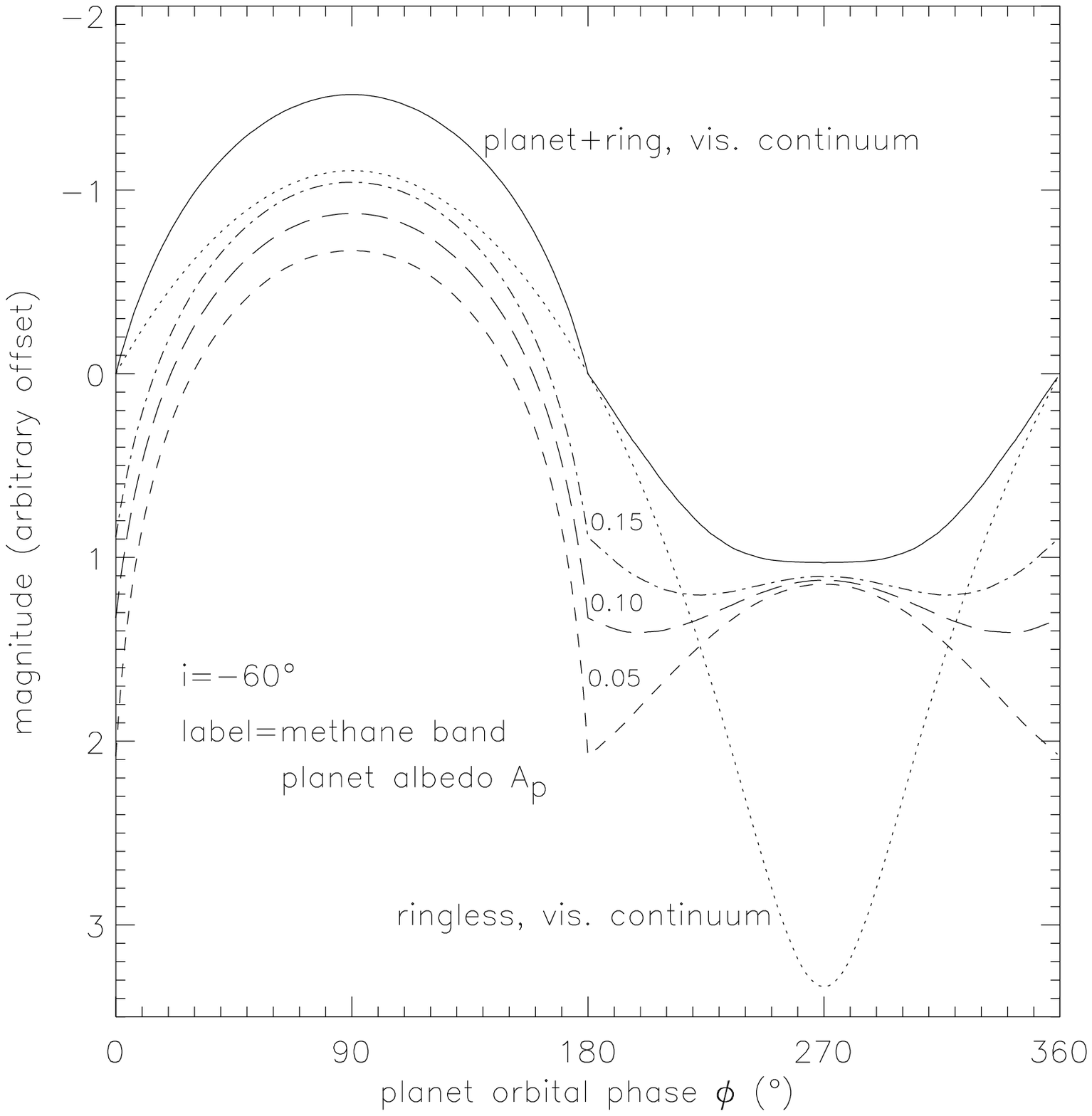}
   \caption{Ringed planet for different orbital positions (labels). Planet obliquity is $5^{\circ}$.
   The pictures upper left represent the planet in the continuum (albedo: planet 0.34, ring 0.7) and below in a $CH_4$ absorption band (albedo: planet 0.05, ring 0.7).
   The corresponding light curves are shown at right, in the continuum and in $CH_4$ absorption band for three different planet albedo. 
   Light curves have significantly different shapes, especially during the second half-orbit: The ring and planet respective brightnesses vary in opposite directions and by the same order of magnitude, making the light curve almost flat. }
   \label{fig02}
\end{figure}

The planet and the ring may have different chemical compositions, and dual-band photometry (or spectroscopy
if the object is bright enough) greatly helps to detect the ring by observing for instance in the (visible) continuum and in
a methane absorption band, where a Saturn-like planet becomes much fainter than the ring. The light curves at both wavelengths can be significantly different, as shown in figure \ref{fig02}. In the methane band, stronger slope changes occur at the equinoxes, here again due to the two observation regimes of the ring, either seen
reflecting or transmitting the light. Note that when the ring disappears at the equinoxes, the object spectrum may be the spectrum of the planet only, but it can also be composite if a part of the planet is seen through the unilluminated ring.

\subsection{The effect of the longitude of ring obliquity}
\begin{figure}
   \includegraphics[width=6.7cm]{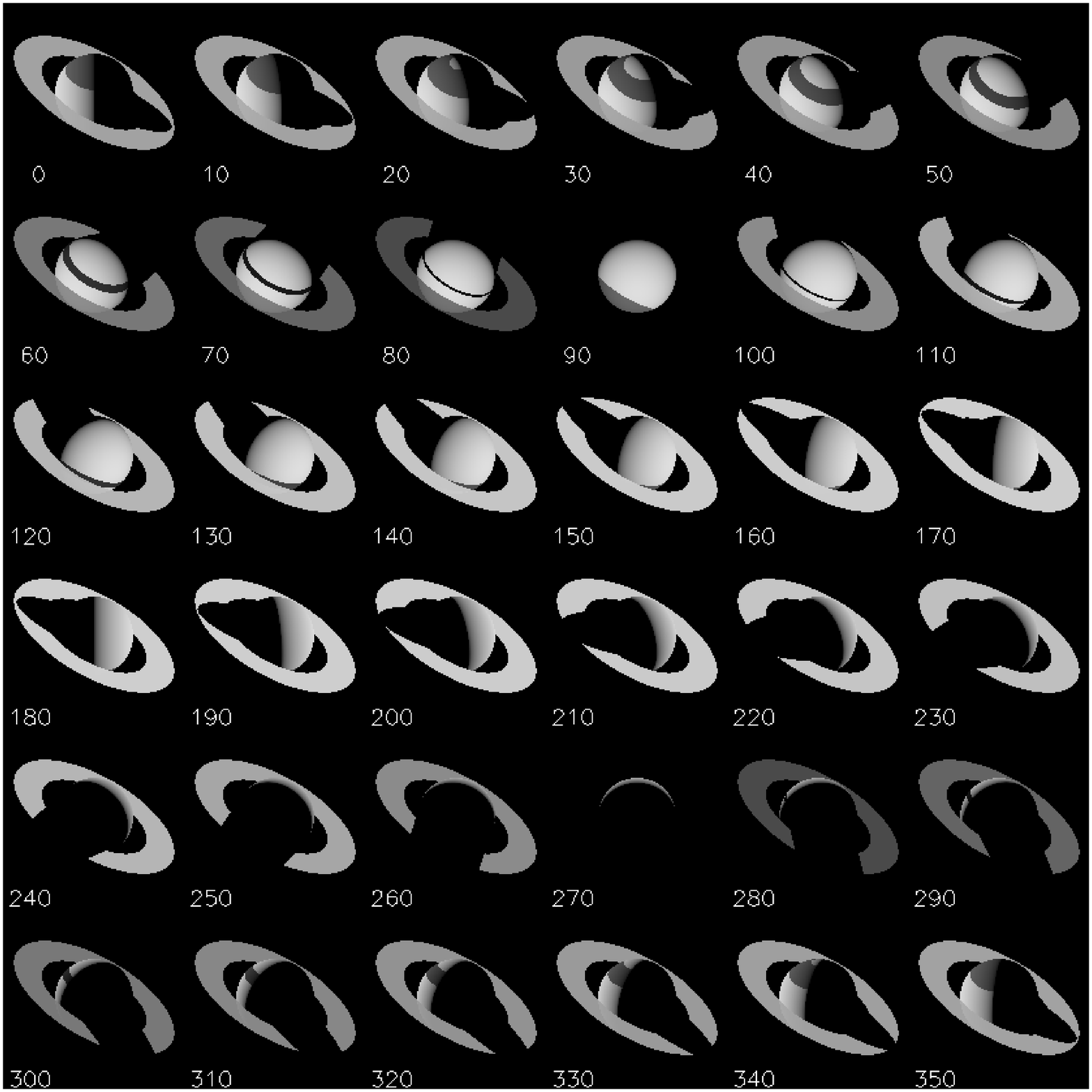}
   \includegraphics[width=6.7cm]{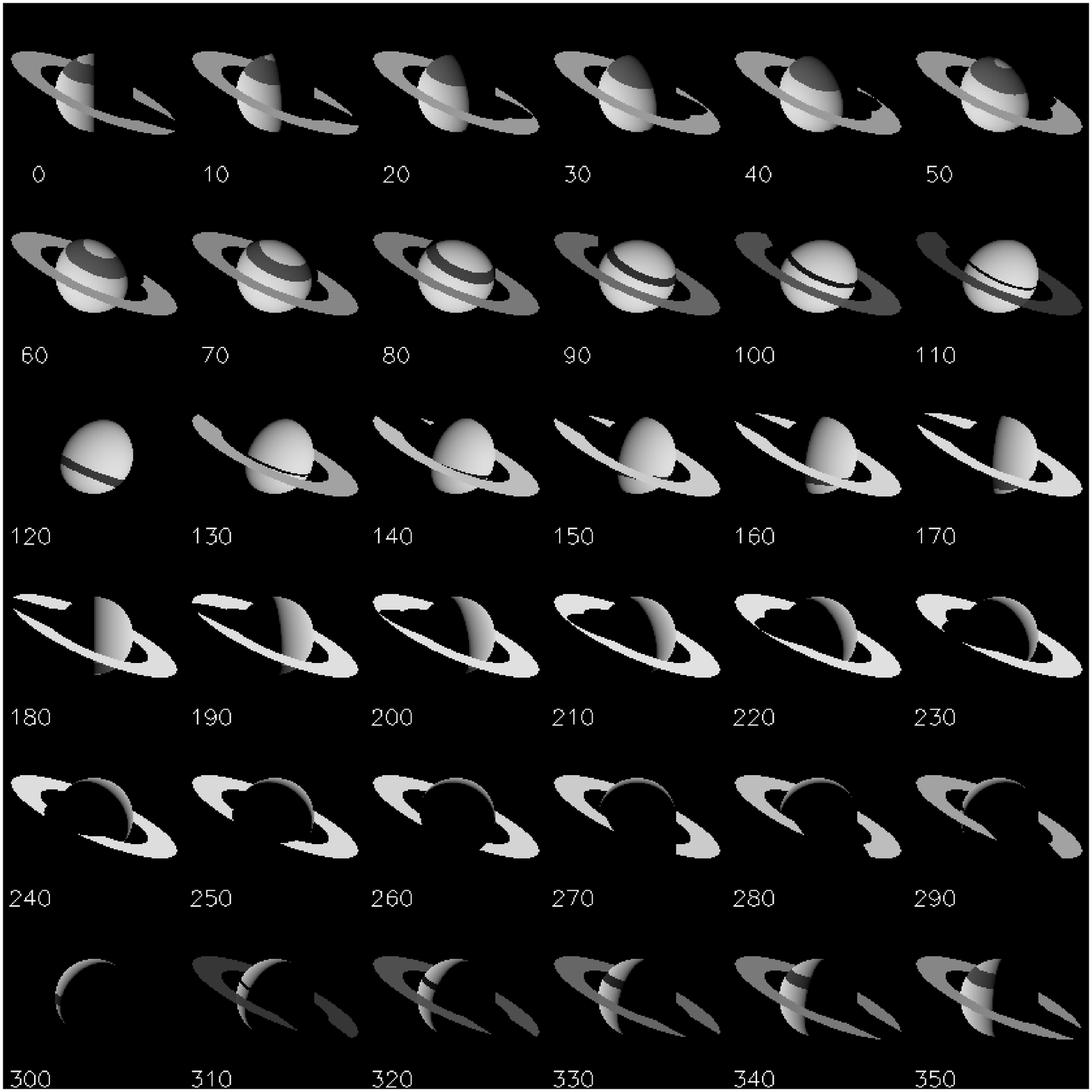}
   \includegraphics[width=6.7cm]{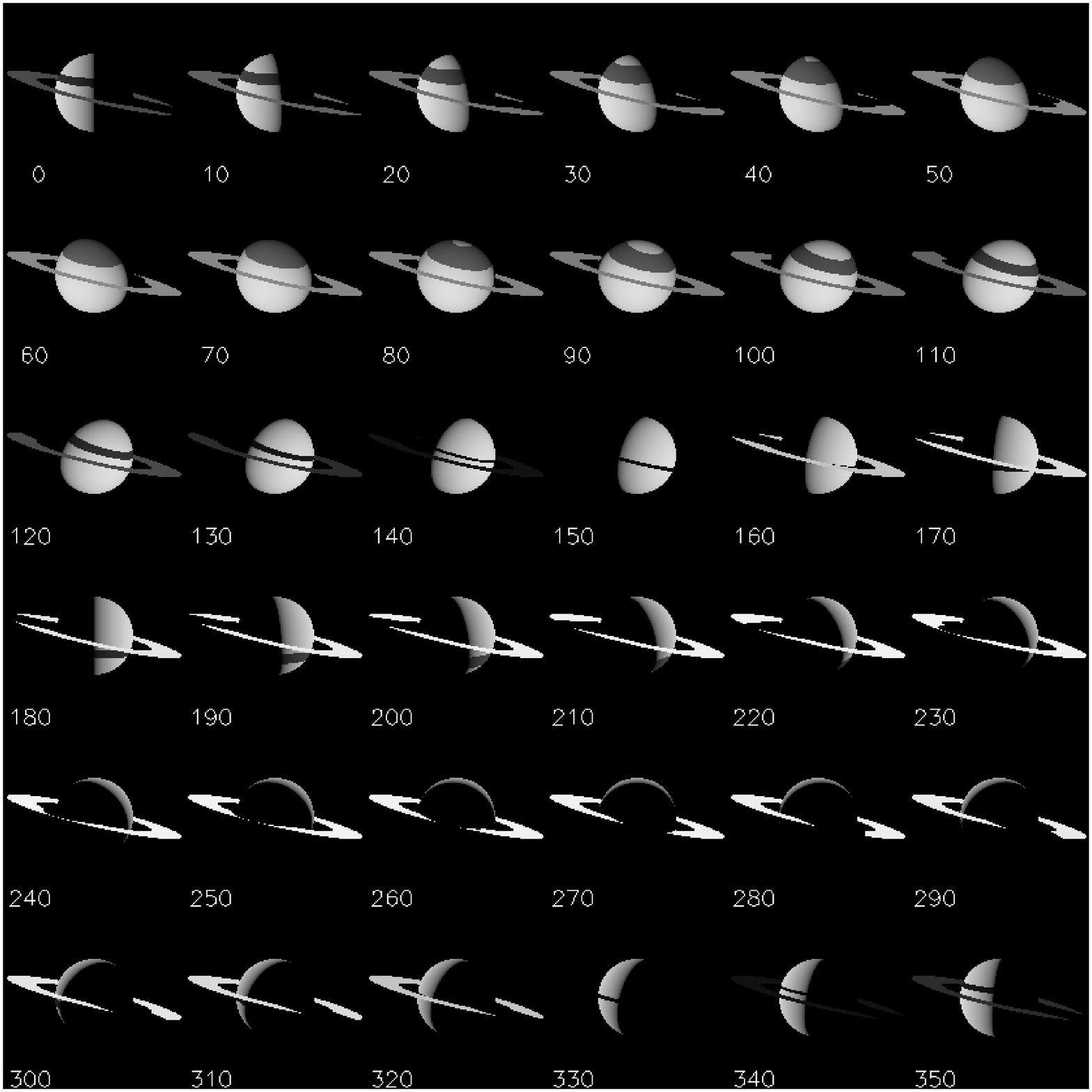}
   \includegraphics[width=6.7cm]{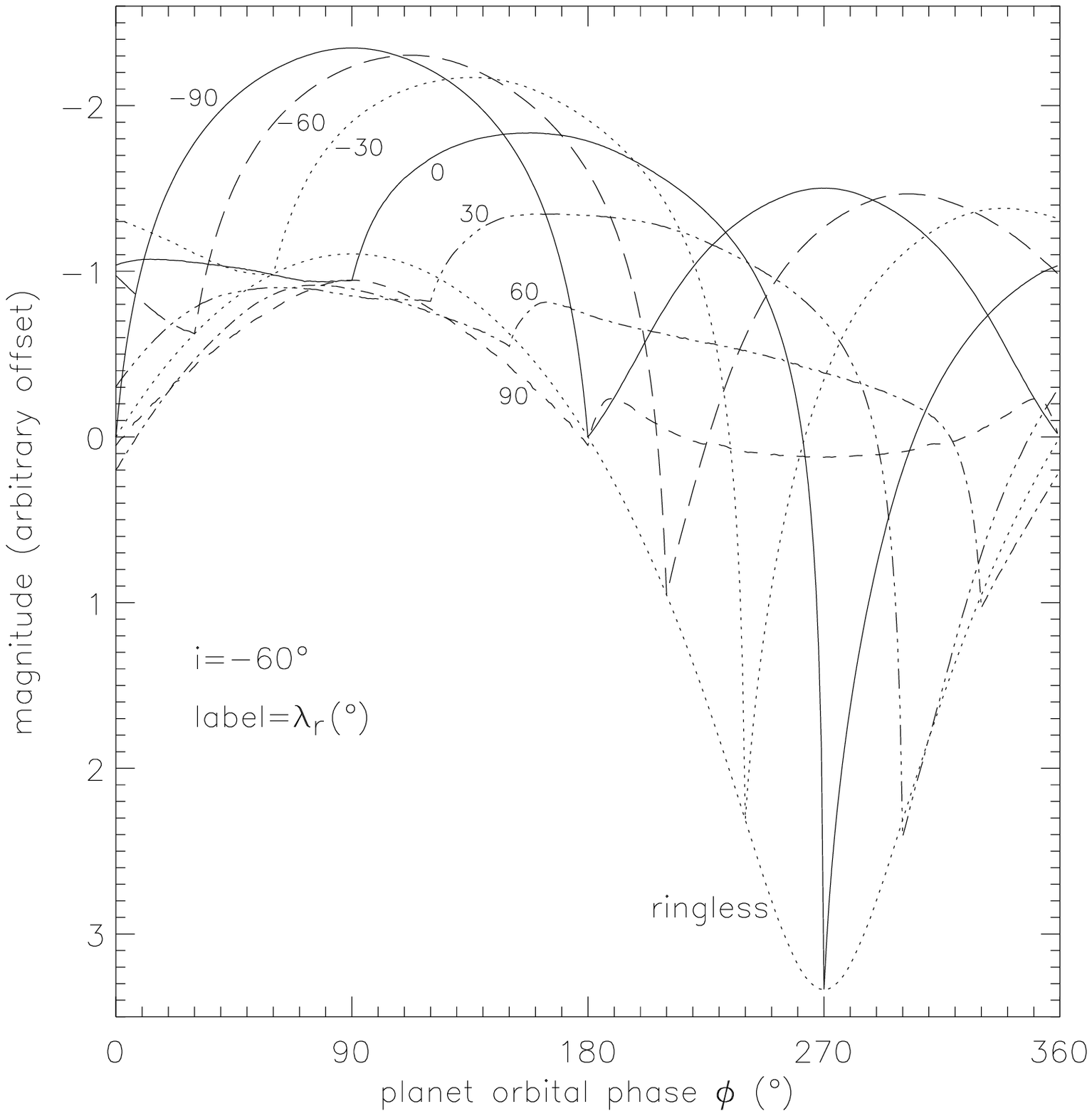}
   \caption{A Saturn-like planet for different orbital positions (labels) and different longitudes $\lambda_r$ of the ring obliquity:
   $\lambda_r=0^{\circ}$ (above left), $\lambda_r=30^{\circ}$ (above right),
   $\lambda_r=60^{\circ}$ (below left). The three sets of images have the same brightness scale.
   These figures show that respective ring and planet brightness extrema do
   not always occur simultaneously. For instance, when $\lambda_r=30^{\circ}$ (above right), a ring minimum occurs at $\phi=300^{\circ}$, while the planet minimum occurs at 		 $\phi=270^{\circ}$. This difference induces a shift of the system light curve as illustrated at lower right.
   The curve for a ringless planet is plotted for comparison. 
   Curves are plotted for $-90^{\circ}\leq\lambda_r\leq90^{\circ}$.
   Note that we have the relation: magnitude$_{\lambda_r}(\phi)=$ magnitude$_{180^{\circ}-{\lambda_r}}(180^{\circ}-\phi)$, all other parameters being unchanged.}
   \label{fig03}
\end{figure}

The orbital angles for which
 the brightness extrema occur do not always correspond with those of a ringless planet. It depends on the ring obliquity longitude.
It is illustrated in figure \ref{fig03}. We consider this {\it$\phi$-shift} of the light curve a relevant signature of a ring. It is unambiguous for a ringed planet on a circular orbit. A $\phi$-shift can be observed if the orbit is elliptical, whether the planet has a ring or not, when brightness variations are induced by distance changes from the planet to its star. But we assume that the astrometry of the discovered planet will make possible the correction of the measured reflected light photometry for the effect of orbit ellipticity.

\subsection{A planet with a large ring}
\begin{figure}
   \includegraphics[width=6.7cm]{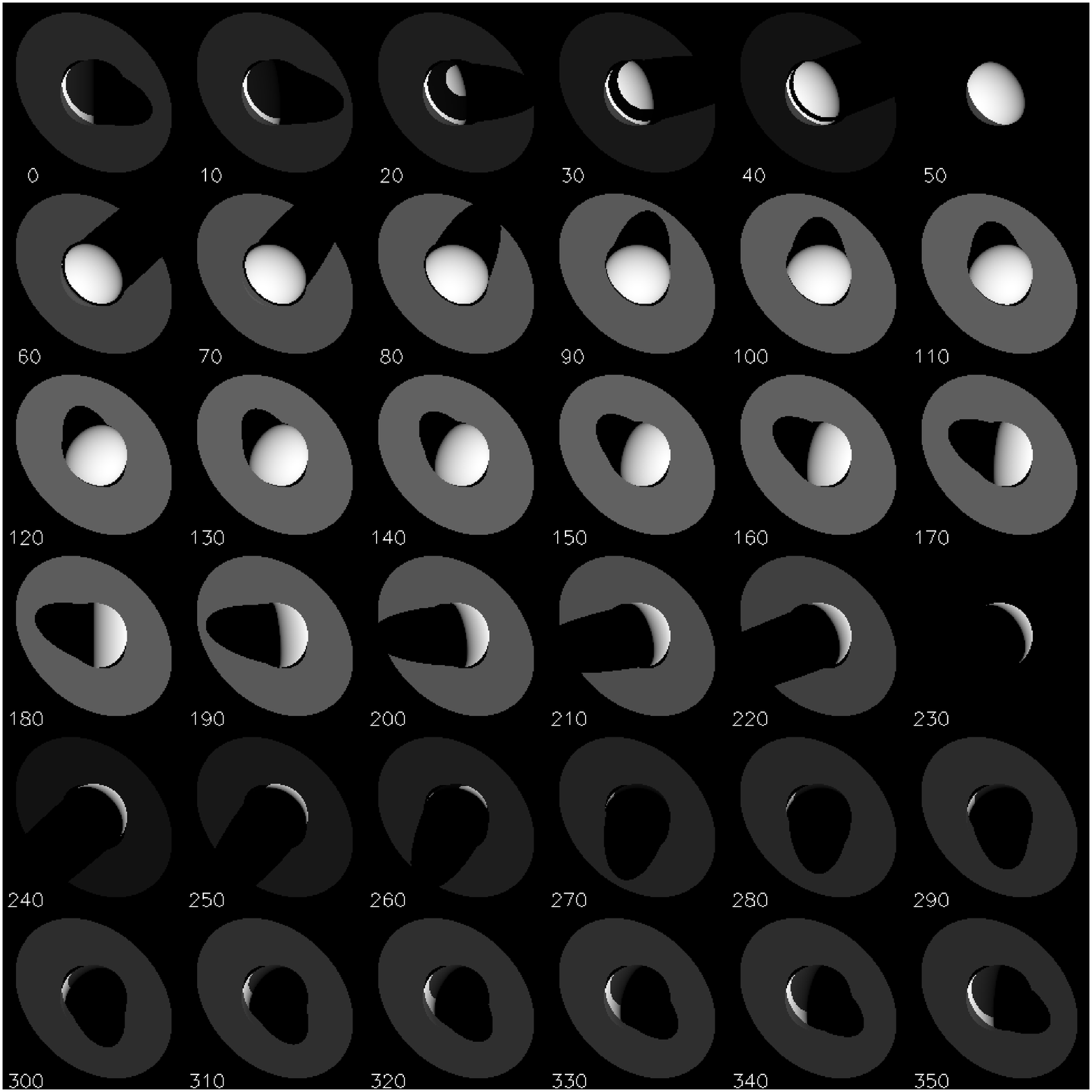}
   \includegraphics[width=6.7cm]{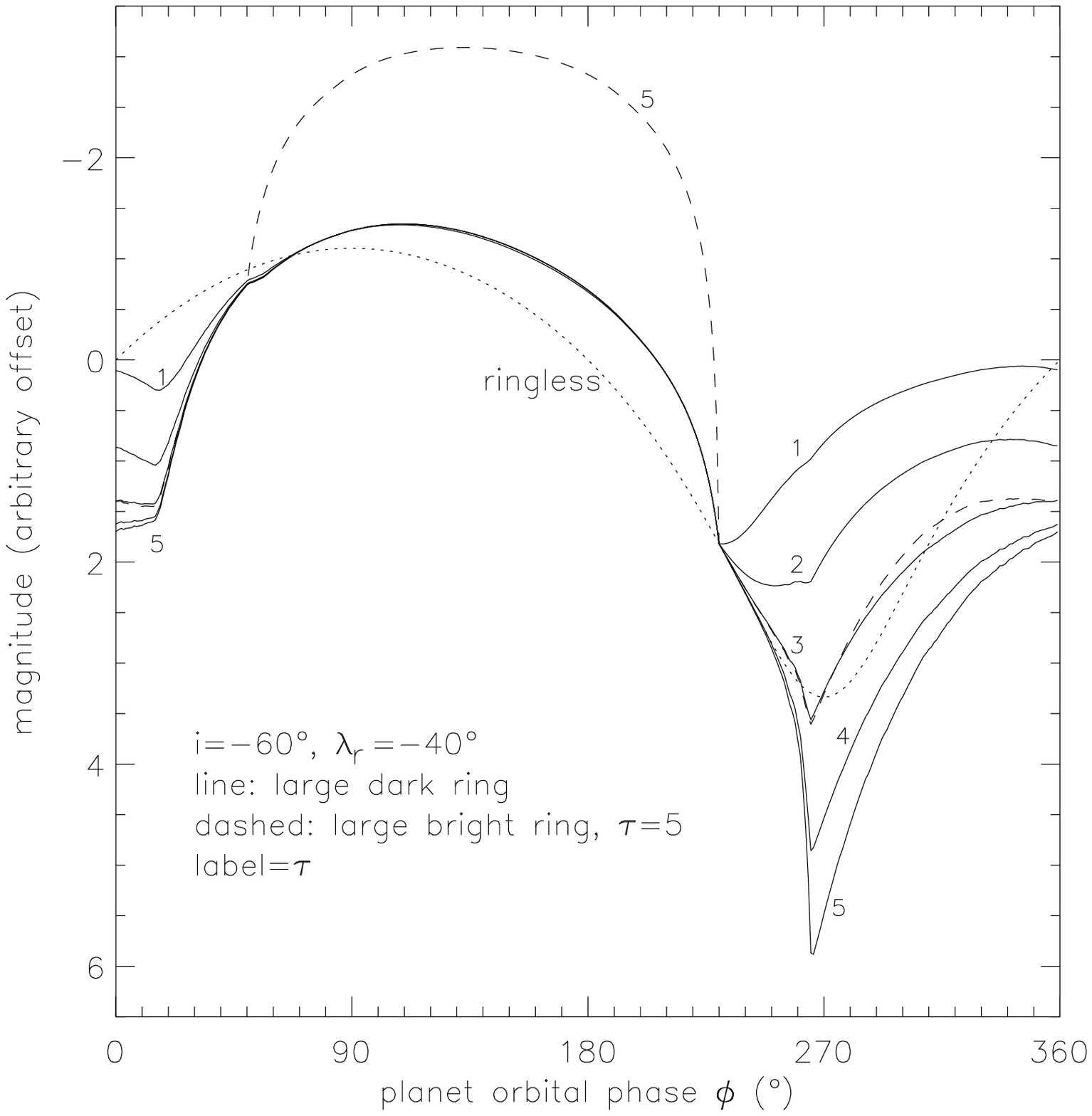}
   \caption{A thick, dark and large ring ($\tau=3$ and albedo $\varpi_0=0.05$) around a planet, seen for different orbital positions $\phi$ (labels). The ring is so large that it hides the planet around $\phi=270^{\circ}$. A curve for a planet with a large, bright and thick ring (albedo $\varpi_0=0.7$ and $\tau=5$) is shown for comparison.}
   \label{fig04}
\end{figure}

For a given planet obliquity, the smaller the gap between the inner edge of the ring and the planet equator, the lower the latitude of the shadow on the planet.
Also, the larger the ring outer diameter, the longer the polar region remains in the shadow of the ring. The ring thus can almost totally hide the planet, as shown
by figure \ref{fig04} around orbital position $\phi=270^{\circ}$. Note that when the planet is hidden, the object spectrum is dominated by the spectrum of the ring.
We consider that such a strong and quite long extinction occurring at phase angles $\alpha>90^{\circ}$ constitutes another specific signature of a ring around an extrasolar planet. It was qualitatively discussed by Schneider (in \cite{desmarais_et_al2002a}).

Of course, the larger (and/or brighter) the ring with respect to the planet, the more the ring dominates the light curve.
But the ratio of the reflected light from the planet over
the stellar flux gives a value of the planetary radius (\cite{schneider01}, 2002). The ring significant flux contribution can 
lead to an overestimation of the planet radius by a factor of 10. If the mass of the planet is known (by radial velocity for instance), the 
overestimated planet radius would lead to an underestimated planet density. Thus a planet with a very low apparent 
density could be a signature of a ring.
Moreover, the planet radius can in principle also be deduced from its infrared thermal flux $F_{p,IR}$, assuming the planet temperature 
has been inferred from the matching of a Planck function to the observed thermal spectrum (\cite{schneider01}, 2002). The disagreement 
between the planet radii at visible and infrared wavelengths could thus be another signature of a ring.

\section{Conclusions}
The examples shown demonstrate that a ring around an extrasolar planet significantly affects the reflected light curve of an extrasolar
planet during its orbital motion. A ring could thus be detected, although both planet and ring would obviously remain unresolved. This may
be achieved simply by a photometric monitoring of the planet light. We identified the following signatures, which would require only
moderate photometric accuracy to be observed (around $3\sigma\approx0.5$ mag):

i) {\it Light curve dichotomy, with strong slope changes at the equinoxes}, due to the ring being alternately seen in reflection and in transmission.

ii) {\it Light curve dependence on wavelength} in dual-band photometry (methane band and continuum for instance), or
{\it spectral variations} if spectroscopy is possible.

iii) The {\it $\phi$-shift} of the light curve extrema, due to the longitude of the ring obliquity.

iv) {\it Temporary extinction of the planet} during the orbital motion, due to the rising shadow of the ring on the planet.

v) {\it High brightness in the reflected light}, leading to an abnormally large planet radius, or/and an abnormally low mass density.

vi) {\it Disagreement between the planet radii} measured from reflected light and thermal infrared emission.

Although future space missions studies concentrate mainly on infrared instruments for technical and scientific
reasons, this work shows the additional interest of shorter wavelengths (visible band) for extrasolar planet characterization.

\end{document}